\documentclass[10pt,aps,prd,superscriptaddress,nofootinbib,nobibnotes,longbibliography,floatfix,twocolumn]{revtex4-2}

\usepackage{graphicx}
\usepackage{dcolumn}
\usepackage{bm}
\usepackage{hyperref}
\usepackage{amsmath}
\usepackage{amssymb}
\usepackage{microtype}
\usepackage{xcolor}
\usepackage{orcidlink}

\newcommand{\pstar}{p^{\ast}}

\usepackage[normalem]{ulem}

\begin{document}

\preprint{Draft v1.0}

\title{Locating critical solutions in numerical relativity using automatic differentiation}

\author{David Bambague \orcidlink{0009-0004-7604-8796}}
\affiliation{%
   Observatorio Astronómico Nacional, Facultad de ciencias, Universidad Nacional de Colombia, Sede Bogotá, Ciudad Universitaria, Bogotá
}

\author{Miguel Bezares \orcidlink{0000-0002-3739-6980}}
\affiliation{Nottingham Centre of Gravity \& School of Mathematical Sciences,
University of Nottingham, University Park, Nottingham NG7 2RD, United Kingdom}

\author{Katy~Clough \orcidlink{0000-0001-8841-1522}}
\affiliation{Centre for Geometry, Analysis and Gravitation, School of Mathematical Sciences, Queen Mary University of London, Mile End Road, London E1 4NS, United Kingdom}

\author{Steven~Tobias \orcidlink{0000-0003-0205-7716}}
\affiliation{School of Physics and Astronomy, The University of Edinburgh, Edinburgh EH9 3FD, UK}

\date{\today}

\begin{abstract}
Numerical relativity is a term used to describe the solution of the full non-linear Einstein Equations as an initial value problem using numerical methods. Individual simulations are computationally expensive and in many applications the high complexity of the parameter space and non-linear sensitivity of the solutions makes locating particular outcomes of interest challenging. In this work, we demonstrate how auto-differentiation can efficiently find dynamical solutions of interest, using the classic problem of scalar critical collapse as a proof of principle. 
We locate a line of critical solutions in a 2D parameter space using 35 runs, compared to the 270 runs required to locate 10 critical points using bisection, and directly calculate the unstable mode, confirming its universality along the line.
The method is straightforwardly extendable to higher-dimensional parameter spaces, offering a promising method to efficiently target more complex solutions in strong dynamical gravity regimes.
\end{abstract}

\maketitle

{\em \textbf{Introduction.}---}
In the spherically symmetric collapse of a matter overdensity in general relativity (GR), there are two possible outcomes. In the first case, the matter overdensity forms a black hole (BH), and in the second, the matter disperses to infinity, leaving behind flat space.
Choptuik \cite{Choptuik:1992jv} first used numerical relativity (NR) simulations to study such cases dynamically, uncovering universal behaviour at the threshold of black-hole formation for the case of a minimally-coupled, massless scalar field.
For any parameter $p$ in the initial data (e.g. amplitude of the initial Gaussian overdensity, or its width), there is a critical value $p_*$ separating subcritical solutions (which disperse) from supercritical solutions (which form black holes). Sufficiently close to the threshold, the mass of the resulting black hole $M_{\rm BH}$ obeys a power-law scaling
\begin{equation}
    M_{\rm BH} \propto \left( p - p^{*}\right)^\gamma ~,
\end{equation}
where $\gamma$ is a universal exponent that does not depend on the parameter $p$ being tuned. (The proportionality constant and the value of $p_*$, on the other hand, depend on the chosen parameter family.)
For massless scalar fields, the value of $\gamma=0.374$ is found numerically by measuring the scaling \cite{Choptuik:1992jv,Garfinkle:1994jb} and by a construction of the critical solution and its perturbations using semi-analytic calculations \cite{Gundlach:1996eg}. The solution exhibits periodic oscillations (``echoes'') with discrete self-similarity - it repeats itself, on a time and length scale that decreases by a factor of $e^{\Delta}$ each time, with $\Delta \approx 3.45$. 
The mass scaling, universality and discrete self-similarity can be understood in dynamical-systems language to mean that the critical solution is an unstable attractor with one unstable eigenvalue. Tuning any parameter closer to the critical value reduces the amplitude of this mode and decreases the unstable eigenvalue, thus allowing the evolution to remain near the critical solution for longer.

Subsequent numerical and semi-analytic studies have confirmed and extended this universality to other matter types in spherical symmetry, but the picture in axisymmetry and beyond remains unclear (see \cite{Gundlach:2007gc} for a review).  Linear perturbation theory indicates that all nonspherical perturbations of the Choptuik solution decay, suggesting that the solution remains a codimension-one attractor in a neighbourhood of spherical symmetry~\cite{Martin-Garcia:1998zqj}. However, fully non-linear simulations revealed additional phenomena~\cite{Choptuik:2003ac,Abrahams:1994nw,Baumgarte:2019fai}. The current status for the vacuum case is that power-law scaling remains, but the critical exponent and self-similar echoing do not appear to be universal across families of initial data \cite{Baumgarte:2023tdh,Baumgarte:2023saw,Fernandez:2022hyx,Mendoza:2021nwq,Ledvinka:2021rve}. 

Like many problems in NR, the study of critical collapse suffers from the expense of individual simulations, and the need to finely tune parameters in a highly non-linear system, in order to achieve a particular evolution. Without prior knowledge, the initial value formulation of NR is essentially a blind search for interesting outcomes. 

Automatic differentiation (AD) is a technique used widely in control theory that provides derivatives of a numerical solution with respect to model parameters without relying on finite differences. It works by applying the chain rule to every operation in the numerical evolution. In forward mode, a tangent state representing the variation of the evolved variables with respect to a chosen parameter is propagated alongside the original state. Consequently, the local change in the solution with respect to that parameter is available at every point in the evolution from a single simulation. AD dates back over fifty years \cite{wengert1964simple,linnainmaa1976taylor,rall1981automatic} with modern implementations rapidly accelerating progress in this area; see e.g. \cite{margossian2019review, maclaurin2015autograd, jax2018github}.
The derivatives obtained in this way are accurate to floating-point precision for the discretised numerical evolution, although they remain subject to the truncation error of the underlying numerical scheme. Propagating the tangent state increases the computational and memory costs for each parameter direction, but the approach is considerably more accurate and efficient than estimating the same sensitivity by performing multiple simulations in an $N$-dimensional space and forming finite differences.

\begin{figure}[t]
  \includegraphics[width=\linewidth]{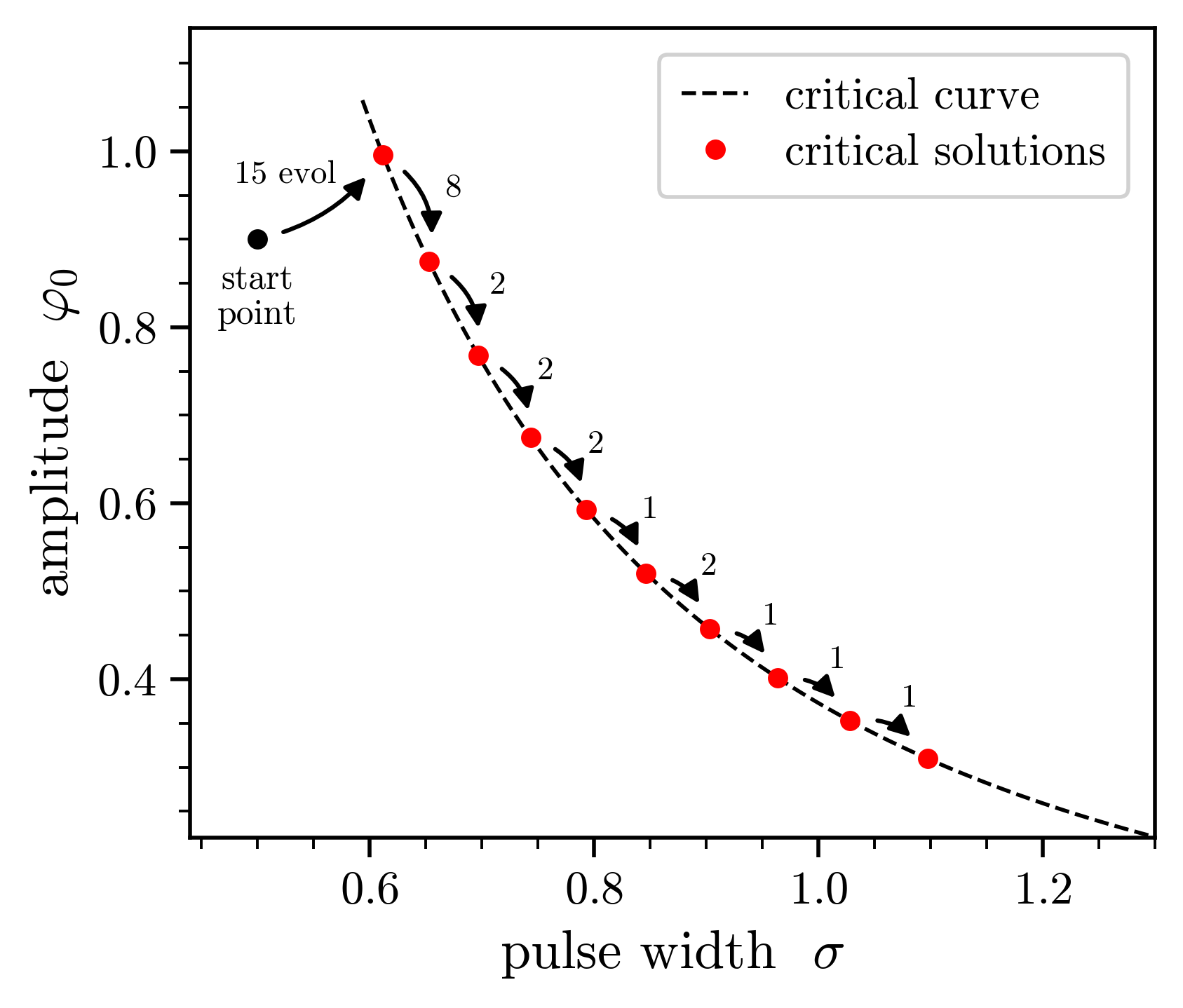}
\caption{\emph{Critical curve of the two-parameter family of initial data $(\sigma,\varphi_0)$.} Red dots show the critical solutions found by the gradient-based search. The black dot is the starting guess, $(\sigma,A)=(0.5,0.9)$. The arrows show the number of evolutions needed to obtain the next critical point on the curve, with 10 points obtained in $35$ evolutions. The dashed line shows the curve interpolated through the critical points, which agrees with that expected from scaling arguments.}
\label{fig:critical_curve}
\end{figure}

In this work, we apply the techniques of AD to the problem of critical collapse, to demonstrate the potential of these methods for studies of strong dynamical gravity spacetimes.
Our goal is to explore the two-dimensional parameter space of width and amplitude of an initial Gaussian perturbation. In this space of initial data, parameters leading to the critical solution lie on a curve, and we can move towards it (from the dispersal side) by increasing the amplitude or decreasing the width of the Gaussian. Once on the line, we can move in the direction that keeps the cost function minimized, thus tracing out the 1D line of initial data that leads to critical solutions, as illustrated in Fig. \ref{fig:critical_curve}.
Choptuik collapse is a classic NR problem, and whilst we restrict to spherical symmetry, we employ a full BSSN + puncture gauge evolution scheme \cite{Ruchlin:2017com,Baumgarte:2015dya} of the type used in standard 3+1D evolutions for gravitational waves. Critical collapse is also a very finely tuned problem requiring high resolutions, and so in this respect is more challenging than other applications in NR.  

This is the first time, to our knowledge, that AD has been applied to a full NR evolution, although it has been used in other contexts such as differentiable surrogate models of numerical-relativity waveforms~\cite{Purrer:2026wgp}, differentiable gravitational-wave waveform modelling and parameter estimation~\cite{Wong:2023lgb,Edwards:2023sak}, and the construction of differentiable binary-black-hole initial data~\cite{DeCeuster:2026pmp}.

{\em \textbf{Methods.}---}
The initial scalar perturbation has the profile $\varphi = \varphi_0\, r^{2}\, e^{-\left(r/\sigma\right)^2}$,
where $\varphi_0$ and $\sigma$ are the amplitude and width of the Gaussian. We obtain constraint-satisfying initial data following the methods in \cite{Werneck:2021kch}, taking a moment of time symmetry (extrinsic curvature $K_{ij}=0$, and $\partial_t\varphi=0$) and a conformally flat spatial metric $\gamma_{ij}=\psi^4\delta_{ij}$, 
so that the momentum constraint is satisfied identically and the Hamiltonian constraint reduces to a linear ODE
for the conformal factor $\psi$,
\begin{equation}
    \frac{1}{r^{2}}\frac{d}{dr}\!\left(r^{2}\frac{d\psi}{dr}\right) = -\pi\,(\partial_r\varphi)^{2}\,\psi~,
\end{equation}
which is solved with boundary conditions $\partial_r\psi|_{r=0}=0$ and $\psi\to1$ as $r\to\infty$. The initial lapse $\alpha$, which determines the proper time that passes for normal observers between hypersurfaces of constant coordinate time $t$, is set to the pre-collapsed profile $\alpha=\psi^{-2}$. Due to rescaling arguments, we know that the initial profile gives critical solutions in the $(\varphi_0, \sigma)$ plane that lie on a curve $\varphi_0^*\sigma_*^2=\mathrm{const}$, and we use this fact to validate our results in Fig. \ref{fig:critical_curve}.

The metric degrees of freedom are evolved using the evolution scheme of \cite{Ruchlin:2017com,Baumgarte:2015dya} implemented in a version of the python-based engrenage code~\cite{engrenage_2026} ported to use the JAX library \cite{jax2018github}, with the equation of motion for the scalar field $g^{\mu\nu}\nabla_\mu\nabla_\nu \varphi = 0$, where $g_{\mu\nu}$ is the metric tensor.

\begin{figure}[t]
  \includegraphics[width=\linewidth]{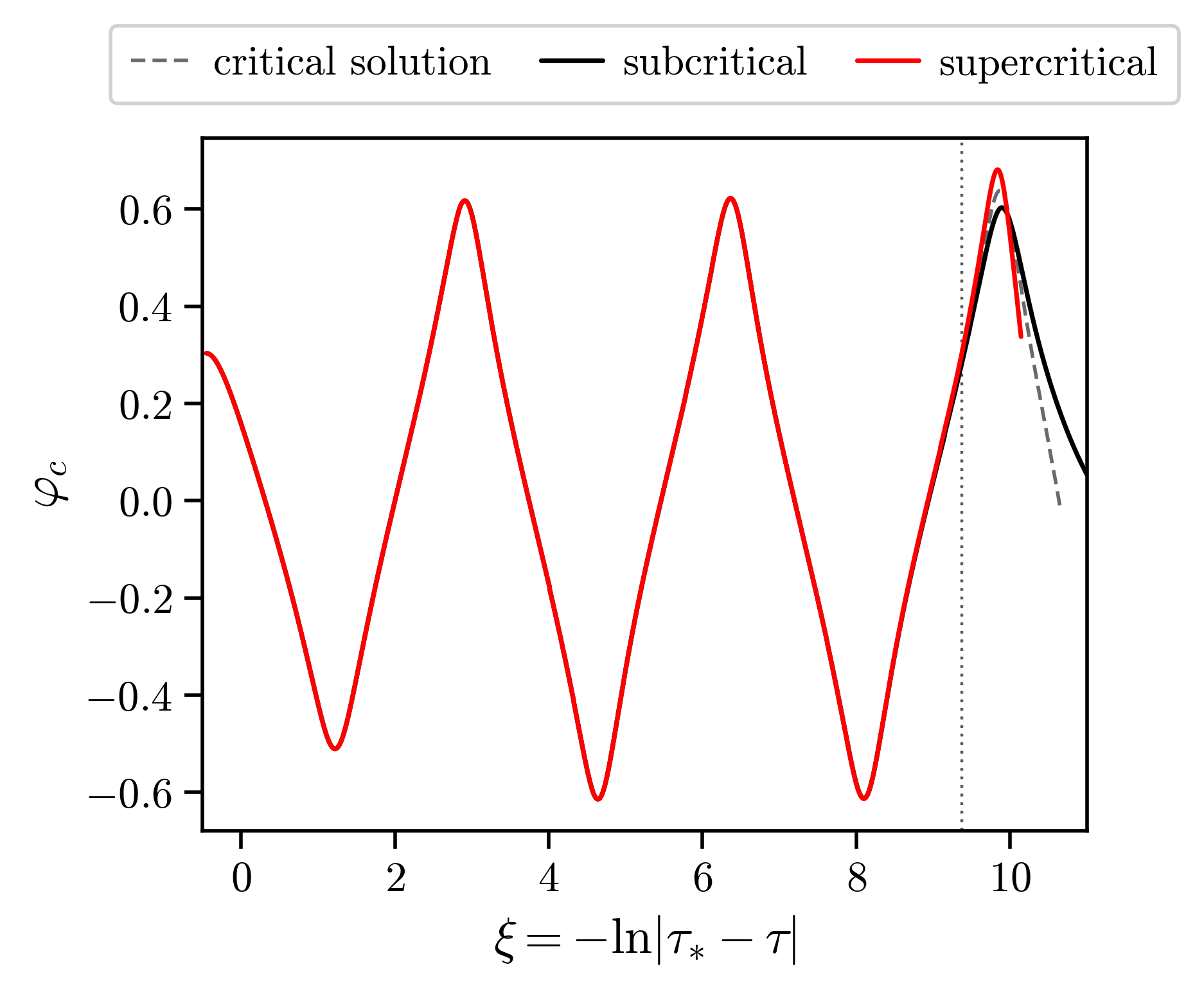}
\caption{\emph{Self-symmetry echoing behaviour.} Central value of the scalar field $\varphi_{c}$ versus $\xi=-\ln|\tau_\ast-\tau|$ (with $\tau$ the central proper time) for a subcritical (black) and a supercritical (red) evolution with $|p-p^\ast|=3.6\times10^{-13}$, together with the run at the best estimate of $p^\ast$ (dashed), used as the critical-solution reference. After the initial transient, the field follows the critical solution with echoing on a period of $\Delta\simeq3.44$. The dotted vertical line marks $\xi\simeq9.4$, where the sub- and supercritical fields first differ by more than $0.02$.}
\label{fig:echoing}
\end{figure}

Each run ends in one of two states - the pulse either disperses (subcritical) or forms a black hole (supercritical). The goal is to tune the parameters $\sigma$ and $\varphi_0$ such that the critical threshold is reached, at which point we observe the typical discrete self-symmetry echoing behaviour, as illustrated in Fig. \ref{fig:echoing}.

The exact threshold is known to depend on the initial discretisation used (since a different discretisation corresponds to a different continuum limit). 
For the initial discretisation used in our work, the threshold is found to be $\pstar = 0.303341761641$ to an accuracy of $10^{-12}$ from a bisection search.
During the evolution, we refine the grid adaptively to follow the structure forming on smaller scales using the methods of \cite{Werneck:2021kch}. However, when neighbouring evolutions are solved on different meshes, the critical amplitude inherits a mesh-dependent scatter at the $10^{-5}$ level. By using a fixed refinement schedule we are able to tune the solution further, to $10^{-9}$.

\begin{figure}
  \includegraphics[width=\linewidth]{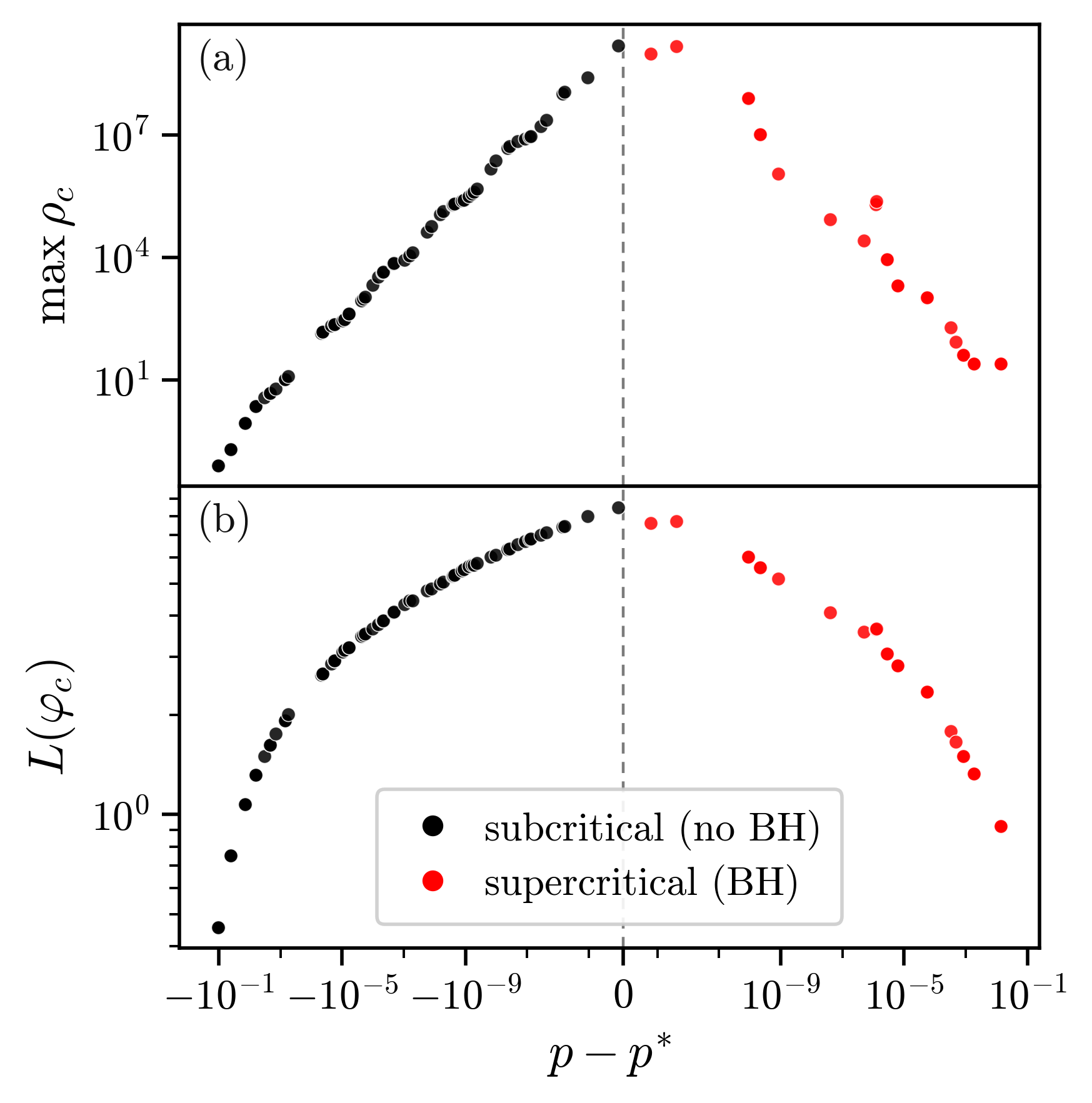}
\caption{\emph{Candidates observables for the (inverse) cost function.} These are measured on 105 evolutions of the Gaussian family with $\sigma=1$, with subcritical (black) and supercritical (red) runs shown as dots. (a) Maximum central energy density $\rho_c$ - as an extreme value attained at a single instant, it is sensitive to the full log-periodic modulation and is unbounded, with a cusp at $p^\ast$. (b) Total variation of the central scalar field, $L(p)$ accumulated along the evolution - it is continuous and differentiable away from $p^\ast$, and because it integrates over all echoes the log-periodic modulation is more suppressed.}
\label{fig:candidate_observables}

\end{figure}

One of the key choices in defining our AD scheme is the choice of {\it cost function} to differentiate. This quantity needs, in general, to be chosen using some physical knowledge of the target solution \cite{Skene2025OnNT}.
As can be seen in Fig. \ref{fig:echoing}, for the closest values on either side of the critical point in the bisection search ($(p-\pstar)\sim \pm 10^{-13}$), the two evolutions are the same through the early echoes and only differ in the final part of the evolution, where the subcritical branch disperses and the supercritical branch collapses. The number of echoes $N$ grows as $\ln|p-\pstar|$, so points further from the critical point follow the same tracks initially but depart from them earlier. As a result of this dynamics, the central energy density $\rho_c$ and Kretschmann scalar $K_c$ grow by orders of magnitude on both sides of the critical point, as a result of the (roughly constant amplitude) echoes in $\varphi_c$ being concentrated in smaller and smaller regions of space and time (see Fig.\ref{fig:echoing}). Evolutions of parameters further from the critical point reach only lower densities and curvatures before they disperse or collapse.
These therefore seem to be obvious candidates for defining the cost function - e.g. the inverse of the maximum of the central density should go to zero in the case of the critical solution. However, we find that quantities that diverge at the critical point do not perform well, partly because of the strong influence of discrete self-similarity features in their value. Other measures were found to suffer from discontinuities or noise in the auto-differentiation. Two candidate cost functions are illustrated in Fig. \ref{fig:candidate_observables}, where we plot the behaviour of the quantity using the results of a bisection search using only variation of the Gaussian amplitude $\varphi_0$.
Both quantities show some superimposed oscillations due to the discrete self-similarity, but these are easiest to see in the quantity $\max\rho_c$. Such oscillations are undesirable because it is generally easier to find the minimum of a smoothly decreasing function. The supercritical points show what appear to be measurement artifacts. These arise partly because a supercritical run is stopped when the central lapse collapses, i.e., $\alpha_c\le5\times10^{-3}$), so its observables are sampled at whatever phase of the last, incomplete echo that value lands on, and this varies with the initial parameters $p$. Additionally, due to the gauge condition used, at some point in the central coordinate point is no longer the physically central point, but maps to a sphere at a finite distance around the singularity. Subcritical runs instead evolve to dispersal, and the central point remains regular, which makes the quantities cleaner to measure.
One could investigate improving this in future by using integrated measures over a spatial region instead of simple central values, although this potentially introduces further gauge dependence.

As illustrated in Fig.~\ref{fig:echoing} and \ref{fig:candidate_observables}, one quantity that we found reasonably smooth is the total variation (integrated arc-length) of the central scalar field value $\varphi_c$, that is
\begin{equation}
  L(p) \;=\; \sum_t \bigl|\Delta\varphi_c(t, p)\bigr| .
\end{equation}
This is accumulated along the whole evolution up to BH formation in the supercritical case, and up to the point when the central lapse value begins to increase monotonically toward unity (breaking the DSS echoes) in the subcritical case.

Since the number of echoes grows like $N\sim\ln|p-\pstar|$, the arc-length inherits the logarithmic law
\begin{equation}
  L(p) \;\simeq\; a + b\,\ln|p-\pstar|, \qquad b<0,
  \label{eq:loglaw}
\end{equation}
where $a$ and $b$ are constants.
Differentiating Eq.~(\ref{eq:loglaw}) gives $L'(p)=b/(p-\pstar)$, so we can estimate the distance to the threshold as
\begin{equation}
  d(p) \;\equiv\; \frac{b}{L'(p)} \;=\; p-\pstar .
  \label{eq:dhat}
\end{equation}
We solve $d(p)=0$ by Newton's method,
\begin{equation}
    p_{k+1}=p_k-\hat d(p_k)=p_k-b/L'(p_k) ~.
\end{equation}
For data that exactly satisfies Eq. (\ref{eq:loglaw}), this would land on $\pstar$ in a single step, but in practice the log-periodic modulation of $L$ limits each step to a reduction of $|p-\pstar|$ by a factor of 10-100. The slope $b$ is estimated at every iteration by linear regression of $L$ against $\ln|L'|$ over the subcritical runs accumulated so far, avoiding the need for knowledge of $\pstar$ using the relation $L=a+b\ln|b|-b\ln|L'|$.
We find $b=-0.268$ for the family of Fig.~\ref{fig:candidate_observables}. 

Each evolution is differentiated by forward-mode AD using the JAX library, which gives the value of $L'(p)$. A Jacobian–vector product (JVP) is propagated through the full nonlinear solution of the PDE, so that one run returns both $L$ and its exact gradient $L'$, from which $d$ is estimated. We find that this results in $\sim 4-5\times$ the cost of a single run for a 2 parameter JVP calculation, but it should be possible to optimise this further. Because $L'$ carries a residual $\sim10\%$ error from the DSS modulation, a step can occasionally overshoot to the other side of the threshold, so the sub/supercritical classification of every run is used to keep a bracket around $\pstar$. We say that we find a solution when $|d|<10^{-9}$ and $\pstar$ is bracketed by a subcritical and a supercritical run within that tolerance.

An alternative to the Newton iteration is a search based on gradient descent, which would be necessary in a higher dimensional parameter space. For example, one could invert Eq.~(\ref{eq:loglaw}) to construct a convex cost, $J=e^{2(L-a)/b}\simeq(p-\pstar)^2$ to be minimised. While we found that this approach broadly worked, it suffered from issues with overshooting the solution and was therefore harder to tune to a high accuracy. We plan to explore this further in a future work.

For two parameters, $\mathbf{x}=(\ln\varphi_0,\ln\sigma)$, Eq. (\ref{eq:loglaw}) holds with $|p-\pstar|$ replaced by the distance to the critical curve, with the gradient $\nabla L$ being the normal to the curve. The Newton step is then
\begin{equation}
    \mathbf{x}_{k+1}= \mathbf{x}_k - b\,\frac{\nabla L}{|\nabla L|^{2}}
                    = \mathbf{x}_k + |d|\,\hat n, \qquad \hat n=\frac{\nabla L}{|\nabla L|}.
\end{equation}

Using this method, our initial search will find an arbitrary point on the critical line in the direction of steepest descent. Further conditions could be applied to specify which point to converge to, e.g. by adding a further penalty objective to the cost function. However, we only care about reaching the line as quickly as possible. Once on the line we can follow the tangent to the curve, using information from previous runs to make the prediction more accurate as we proceed.  

\begin{figure}[t]
  \includegraphics[width=\linewidth]{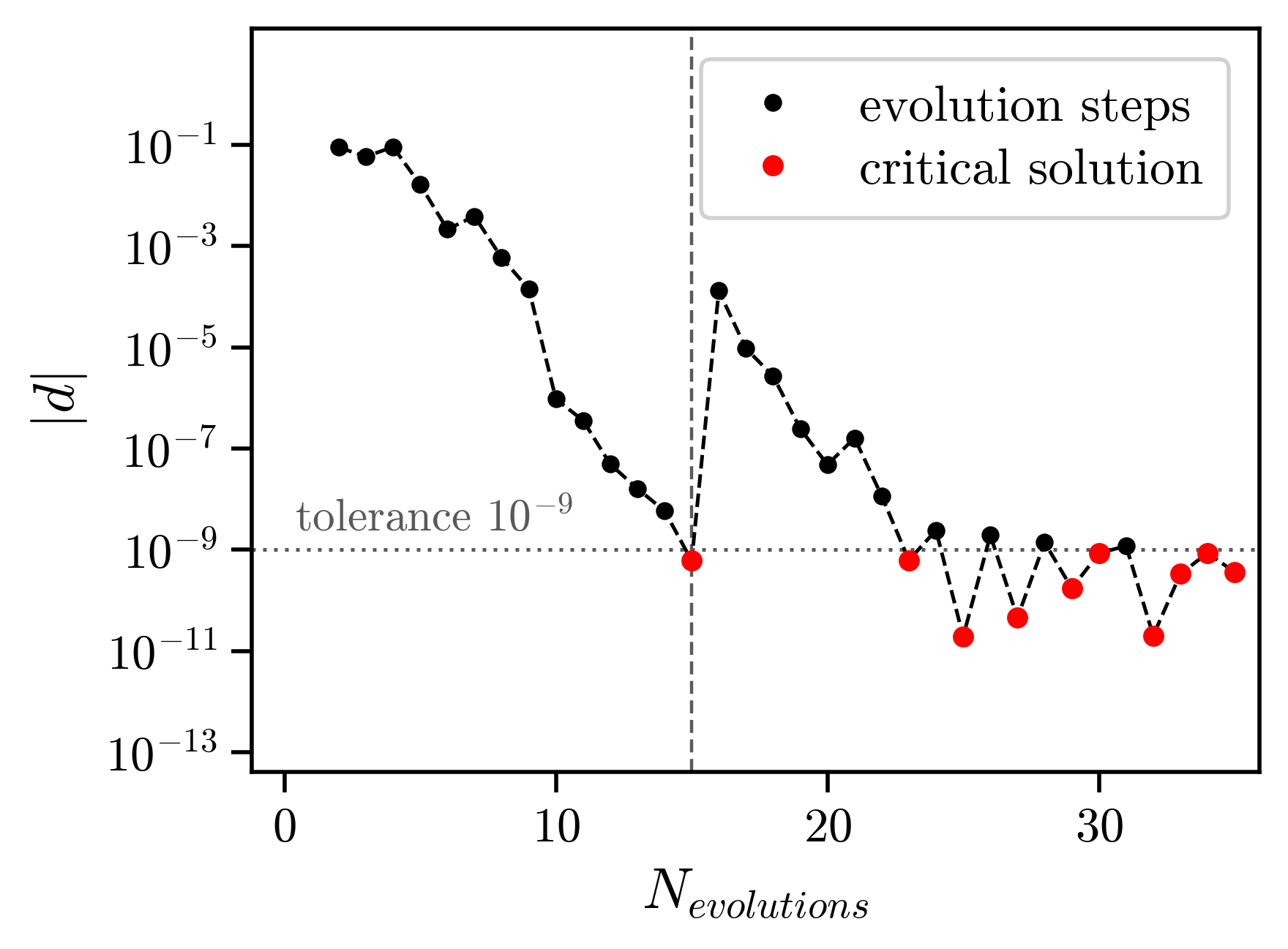}
\caption{\emph{Convergence of the two-parameter search}. Estimated distance to the critical curve, $|d|=|b|/\|\nabla{L}\|$, after each evolution, as a function of the cumulative number of JAX-enabled evolutions $N_{\rm evolutions}$. Black dots are individual evolutions and red dots are those accepted as critical solutions ($|d|<10^{-9}$), shown in Fig.~\ref{fig:critical_curve}. We start at a distance of $|d|\sim0.1$ from which we require 15 evolutions to reach the first critical point (on the dashed line). We then walk along the curve, using information from past evolutions to reduce the number of correction steps required.}
\label{fig:convergence}
\end{figure}

{\em \textbf{Results.}---}
We show our results for the two-parameter search in Fig. \ref{fig:convergence}, illustrating the number of iterations required to reach the solutions in Fig. \ref{fig:critical_curve}. The convergence is slower initially and accelerates as the search refines. This is because far from the threshold the evolution does not closely approach the critical solution, and so the gradient carries less information. The first four evolutions therefore reduce $|d|$ only by a factor of 10, but once the runs start to see the echoing, each step reduces the distance more efficiently (going from $10^{-4}$ to $10^{-9}$ in the 9th to 15th evolutions). Once the first solution is reached, we take a step along the local tangent of the curve, with the tangent and the step length built from the gradients of the previous points. Further Newton iterations are then used to correct this step back onto the critical curve, as shown after the vertical dashed line in Fig. \ref{fig:convergence}. The first such step, with no history, lands at $|d|\simeq10^{-4}$ and needs 8 evolutions to regain the curve, but afterwards the predictor lands close to $|d|\sim10^{-9}$, so every new critical point costs only one or two evolutions, with 20 evolutions used for the nine remaining points.

\begin{figure}[t]
  \includegraphics[width=\linewidth]{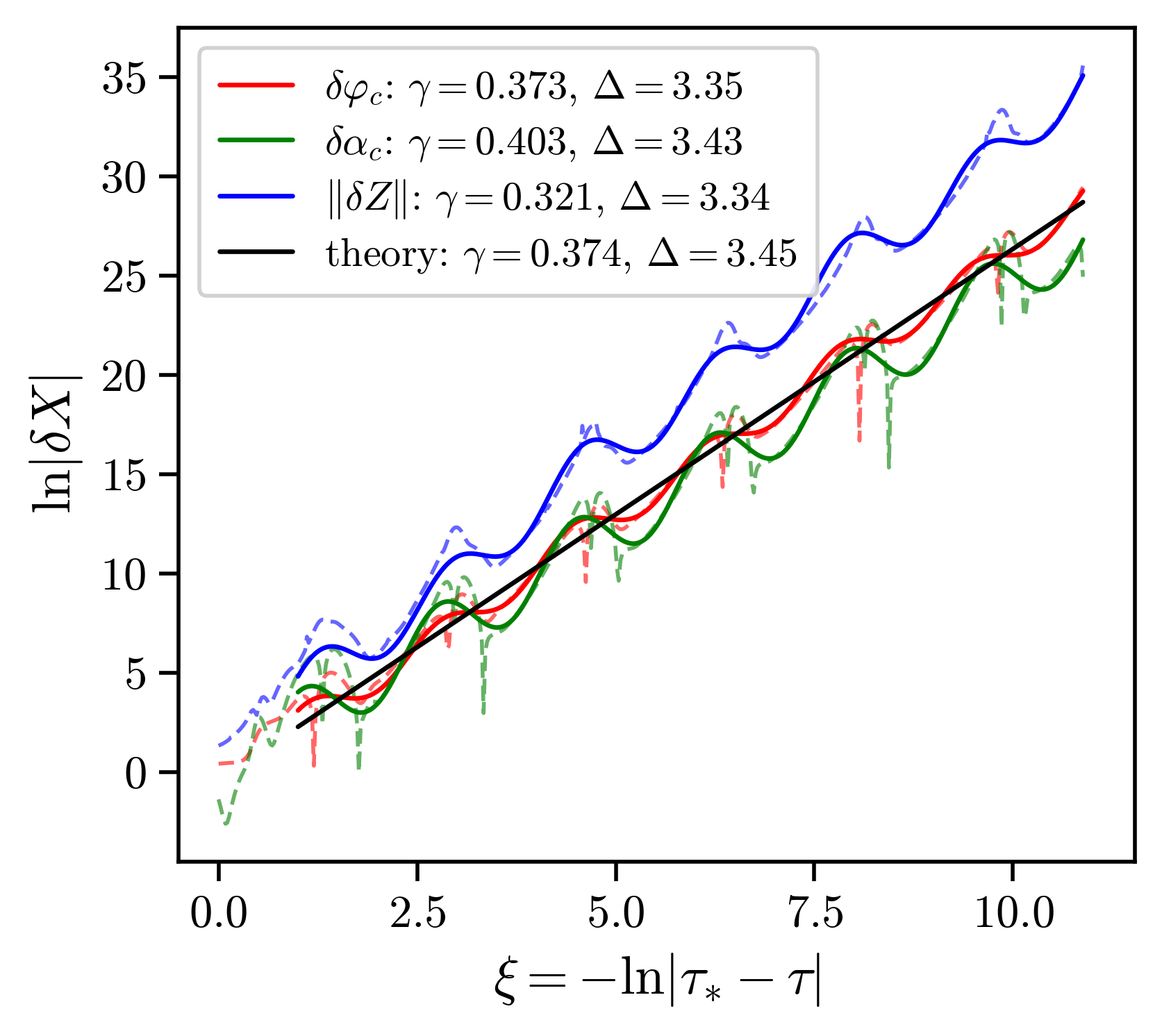}
\caption{\emph{Fitting the variation of the solution}. The variation is shown with respect to three parameters: the central scalar field $\delta\varphi_c$ (red), the central lapse $\delta\alpha_c$ (green), and the norm of the full state vector, $\|\delta Z\|$ (blue). The simulation curves are shown as dotted lines and the fits in solid lines are linear growth plus a sinusoid of period $\Delta$. The inferred value of $\gamma$ for each is given in the legend, to be compared against the known value $\gamma=0.374$ (black line). The central scalar field, as a single gauge-invariant degree of freedom, gives the best result.}
\label{fig:dphi_growth_3obs}
\end{figure}

We can use the AD results to check that the inferred critical exponent remains constant at each critical value, which confirms universality. We consider the growth of the linear perturbation of the central scalar field along the near-critical evolution ($|p-p^\ast|\lesssim10^{-12}$), along with several other evolution quantities, as shown in Fig.\ref{fig:dphi_growth_3obs}.
Here, the quantity $\delta\varphi_c\equiv\partial\varphi(t,r=0)/\partial p$ is not obtained from differences of neighbouring runs, but is the tangent of the full evolution with respect to the amplitude, as obtained from AD. The tangent is renormalised at each step and the accumulated logarithm of the norm is added back, so that $\ln|\delta\varphi_c|$ is followed over $\sim30$ e-folds without numerical overflow occurring. 
This allows us to visualise directly the single unstable mode of the critical solution, whose slope is the Lyapunov exponent of the single unstable mode of the critical solution (its inverse is the critical exponent $\gamma$ \cite{Gundlach:2007gc}). The superposed oscillation is the periodic modulation of the mode by the discretely self-similar background.
We therefore fit a linear growth plus modulations of period $\Delta$ to the observed mode, of the form
\begin{equation}
  \ln|\delta\varphi_c| = c + \kappa\,\xi + k_1\sin\!\left(\frac{2\pi\xi}{\Delta}+\phi_1\right) + k_2\sin\!\left(\frac{4\pi\xi}{\Delta}+\phi_2\right),
\end{equation}
where $c,\kappa,k_1,k_2, \phi_1,\phi_2$ are constants. The fitted slope and period values give $\gamma=1/\kappa=0.3727$ and $\Delta=3.35$, compared to the known values $\gamma=0.374$ and $\Delta=3.445$ \cite{Gundlach:1996eg}, showing agreement at the $0.4\%$ level for $\gamma$ and $3\%$ for $\Delta$. We find that the central scalar value is the variable that gives the best agreement with the expected values. The lapse, for example, depends on the slicing condition, so gauge dynamics adds oscillations and a drift to its tangent, biasing the slope by $8\%$. The norm of the full state $Z$ (the $L^2$ norm of the tangent of all evolved variables) mixes all fields and all radii, including gauge variables and the outer, non-self-similar region, and overestimates the growth rate by $14\%$, as can be seen in Fig. \ref{fig:dphi_growth_3obs}.

{\em \textbf{Discussion.}---}
In this work we have demonstrated as a proof of principle that AD methods can be used to efficiently locate dynamical solutions in numerical relativity simulations with desired properties.

We used the canonical example of scalar field critical collapse, in which any one-parameter family of initial data can be tuned to find the same single critical solution at the threshold of gravitational collapse. We showed that tuning multiple parameters simultaneously with an appropriate cost function gave a faster convergence to a critical solution than a simple bisection method, and then allowed us to map out the line of critical solutions with only a few iterations at each step. While the gains here are relatively modest, the power of the methods is potentially significant for more general cases. In particular, in many cases it is not at all clear in which direction the parameters should be tuned to reach a certain solution (unlike in critical collapse, where one knows to increase the amplitude to find a black hole, or decrease for dispersal). The two parameter space we study can also be simply extended to hundreds of parameters in a conceptually straightforward way, albeit with associated increased costs in memory and runtime.

Our code can also be straightforwardly extended to axisymmetry in the formulation we are using. Introducing AD into full Cartesian 3+1D codes with adaptive mesh refinement is a more significant challenge, but in principle methods exist to do this. For example, AD has been implemented via adjointing both for the finite element Firedrake code \cite{FiredrakeUserManual} and in the Open Source spectral solver Dedalus \cite{skene2026fast}.
The implementation of our JAX enabled code will be made open source shortly, allowing others to apply it to different examples. The main difficulty in applying the AD approach to new scenarios is in selecting the relevant cost function. Here we used prior knowledge about the echoing behaviour near criticality to define an appropriate measure. Even in this case, some amount of trial and error was required to find a useful quantity.

This approach could have applications in many areas of NR. The most obvious is in extending studies of critical collapse beyond spherical symmetry to axisymmetry, where recent studies have indicated that the critical solution is no longer unique \cite{Baumgarte:2023tdh}, and we are currently applying it to identify initial data leading to particular black hole solutions, such as extremal ones \cite{Crump:2026kgu,East:2025nfb}. One could, in principle, use such methods to directly reconstruct the intrinsic black hole parameters leading to a specific observed strain waveform, in cases such as GW231123 where waveform models appear to suffer from systematic disagreements \cite{LIGOScientific:2025rsn}. These are just a few obvious examples, but the overall idea - to convert blind searches in the initial value formulation of NR into targeted searches for specific dynamical outcomes - is applicable to a wide range of studies and could be revolutionary for the field.

{\em \textbf{Acknowledgements.}---}
We would like to thank Josu Aurrekoetxea, Thomas Baumgarte, Daniela Cors, Thomas Helfer and Jorge Santos for helpful discussions, and the GRTL Collaboration (\url{www.grtlcollaboration.org}) for their development work on the engrenage code. 

SMT and KC are grateful to the Higgs Centre for support at the 5th “New Directions in Theoretical Physics” meeting in 2026, where this project was first conceived.

DB is supported by a Teaching Assistantship Scholarship (BAD) from the Observatorio Astronómico Nacional, Universidad Nacional de Colombia.

MB acknowledges partial support from the STFC Consolidated Grant nos. ST/Z000424/1, UKRI2492 and ST/C004235/1.

KC is supported by an STFC Ernest Rutherford fellowship, project reference ST/V003240/1, an STFC Research Grant ST/X000931/1 (Astronomy at Queen Mary 2023-2026), and the Simons Foundation International and the Simons Foundation through Simons Foundation grant SFI-MPS-BH-00012593-13. 

For the purpose of Open Access, the author has applied a CC BY public copyright licence to any Author Accepted Manuscript version arising from this submission.

\bibliography{mybib}

@article{Choptuik:2003ac,
    author = "Choptuik, Matthew W. and Hirschmann, Eric W. and Liebling, Steven L. and Pretorius, Frans",
    title = "{Critical collapse of the massless scalar field in axisymmetry}",
    eprint = "gr-qc/0305003",
    archivePrefix = "arXiv",
    doi = "10.1103/PhysRevD.68.044007",
    journal = "Phys. Rev. D",
    volume = "68",
    pages = "044007",
    year = "2003"
}

@article{Abrahams:1994nw,
    author = "Abrahams, A. M. and Evans, C. R.",
    title = "{Universality in axisymmetric vacuum collapse}",
    doi = "10.1103/PhysRevD.49.3998",
    journal = "Phys. Rev. D",
    volume = "49",
    pages = "3998--4003",
    year = "1994"
}

@article{Martin-Garcia:1998zqj,
    author = "Martin-Garcia, Jose M. and Gundlach, Carsten",
    title = "{All nonspherical perturbations of the Choptuik space-time decay}",
    eprint = "gr-qc/9809059",
    archivePrefix = "arXiv",
    doi = "10.1103/PhysRevD.59.064031",
    journal = "Phys. Rev. D",
    volume = "59",
    pages = "064031",
    year = "1999"
}

@misc{Purrer:2026wgp,
    author = {P{\"u}rrer, Michael and Girish, Ashwin and Thomas, Lucy M. and Field, Scott E. and Varma, Vijay},
    title = "{Fast, accurate, and differentiable: a neural-network surrogate for NRSur7dq4 precessing binary black hole waveforms}",
    eprint = "2607.24960",
    archivePrefix = "arXiv",
    primaryClass = "gr-qc",
    year = "2026"
}

@article{Edwards:2023sak,
    author = "Edwards, Thomas D. P. and Wong, Kaze W. K. and Lam, Kelvin K. H. and Coogan, Adam and Foreman-Mackey, Daniel and Isi, Maximiliano and Zimmerman, Aaron",
    title = "{Differentiable and hardware-accelerated waveforms for gravitational wave data analysis}",
    eprint = "2302.05329",
    archivePrefix = "arXiv",
    primaryClass = "astro-ph.IM",
    doi = "10.1103/PhysRevD.110.064028",
    journal = "Phys. Rev. D",
    volume = "110",
    number = "6",
    pages = "064028",
    year = "2024"
}

@article{Wong:2023lgb,
    author = "Wong, Kaze W. K. and Isi, Maximiliano and Edwards, Thomas D. P.",
    title = "{Fast Gravitational-wave Parameter Estimation without Compromises}",
    eprint = "2302.05333",
    archivePrefix = "arXiv",
    primaryClass = "astro-ph.IM",
    doi = "10.3847/1538-4357/acf5cd",
    journal = "Astrophys. J.",
    volume = "958",
    number = "2",
    pages = "129",
    year = "2023"
}

@misc{DeCeuster:2026pmp,
    author = "De Ceuster, Frederik and Li, Tjonnie G. -F.",
    title = "{A Differentiable Parametric Model of Binary-Black-Hole Initial Data: I. Conformally flat Bowen-York punctures}",
    eprint = "2609.10173",
    archivePrefix = "arXiv",
    primaryClass = "gr-qc",
    year = "2026"
}

@article{wengert1964simple,
  title = {A Simple Automatic Derivative Evaluation Program},
  author = {Wengert, Robert Edwin},
  journal = {Communications of the ACM},
  volume = {7},
  number = {8},
  pages = {463--464},
  year = {1964},
  doi = {10.1145/355586.364791}
}

@article{Skene2025OnNT,
  author    = {Skene, Calum S. and Marcotte, Florence and Tobias, Steven M.},
  title     = {On nonlinear transitions, minimal seeds and exact solutions for the geodynamo},
  journal   = {Journal of Fluid Mechanics},
  year      = {2025},
  volume    = {1021},
  pages     = {1--28},
  articleno = {A37},
  doi       = {10.1017/jfm.2025.10639}
}

@article{linnainmaa1976taylor,
  title = {Taylor Expansion of the Accumulated Rounding Error},
  author = {Linnainmaa, Seppo},
  journal = {BIT Numerical Mathematics},
  volume = {16},
  number = {2},
  pages = {146--160},
  year = {1976},
  doi = {10.1007/BF01931367}
}

@book{rall1981automatic,
  title = {Automatic Differentiation: Techniques and Applications},
  author = {Rall, Louis B.},
  series = {Lecture Notes in Computer Science},
  volume = {120},
  publisher = {Springer},
  address = {Berlin, Heidelberg},
  year = {1981},
  doi = {10.1007/3-540-10861-0}
}

@article{margossian2019review,
  title = {A Review of Automatic Differentiation and its Efficient Implementation},
  author = {Margossian, Charles C.},
  journal = {WIREs Data Mining and Knowledge Discovery},
  volume = {9},
  number = {4},
  pages = {e1305},
  year = {2019},
  doi = {10.1002/widm.1305}
}

@misc{maclaurin2015autograd,
  title = {Autograd: Effortless Gradients in {NumPy}},
  author = {Maclaurin, Dougal and Duvenaud, David and Adams, Ryan P.},
  year = {2015},
  howpublished = {ICML 2015 AutoML Workshop},
  url = {https://github.com/HIPS/autograd}
}

@misc{jax2018github,
  title = {{JAX}: Composable Transformations of {Python}+{NumPy} Programs},
  author = {Bradbury, James and Frostig, Roy and Hawkins, Peter and Johnson, Matthew James and Leary, Chris and Maclaurin, Dougal and Necula, George and Paszke, Adam and VanderPlas, Jake and Wanderman-Milne, Skye and Zhang, Qiao},
  year = {2018},
  howpublished = {\url{https://github.com/google/jax}}
}

@manual{FiredrakeUserManual,
  title        = {Firedrake User Manual},
  author       = {David A. Ham and Paul H. J. Kelly and Lawrence Mitchell and Colin J. Cotter and Robert C. Kirby and Koki Sagiyama and Nacime Bouziani and Sophia Vorderwuelbecke and Thomas J. Gregory and Jack Betteridge and Daniel R. Shapero and Reuben W. Nixon-Hill and Connor J. Ward and Patrick E. Farrell and Pablo D. Brubeck and India Marsden and Thomas H. Gibson and Miklós Homolya and Tianjiao Sun and Andrew T. T. McRae and Fabio Luporini and Alastair Gregory and Michael Lange and Simon W. Funke and Florian Rathgeber and Gheorghe-Teodor Bercea and Graham R. Markall},
  organization = {Imperial College London and University of Oxford and Baylor University and University of Washington},
  edition      = {First edition},
  year         = {2023},
  month        = {5},
  doi          = {10.25561/104839},
}

@article{skene2026fast,
  author    = {Skene, Calum S. and Burns, Keaton J.},
  title     = {Fast automated adjoints for spectral PDE solvers},
  journal   = {Proceedings of the National Academy of Sciences},
  volume    = {123},
  number    = {15},
  pages     = {e2530440123},
  year      = {2026},
  doi       = {10.1073/pnas.2530440123},
  url       = {https://doi.org/10.1073/pnas.2530440123}
}

@article{Choptuik:1992jv,
    author = "Choptuik, Matthew W.",
    title = "{Universality and scaling in gravitational collapse of a massless scalar field}",
    reportNumber = "FPRINT-92-33",
    doi = "10.1103/PhysRevLett.70.9",
    journal = "Phys. Rev. Lett.",
    volume = "70",
    pages = "9--12",
    year = "1993"
}

@article{Gundlach:2007gc,
    author = "Gundlach, Carsten and Martin-Garcia, Jose M.",
    title = "{Critical phenomena in gravitational collapse}",
    eprint = "0711.4620",
    archivePrefix = "arXiv",
    primaryClass = "gr-qc",
    doi = "10.12942/lrr-2007-5",
    journal = "Living Rev. Rel.",
    volume = "10",
    pages = "5",
    year = "2007"
}

@article{Garfinkle:1994jb,
    author = "Garfinkle, David",
    title = "{Choptuik scaling in null coordinates}",
    eprint = "gr-qc/9412008",
    archivePrefix = "arXiv",
    doi = "10.1103/PhysRevD.51.5558",
    journal = "Phys. Rev. D",
    volume = "51",
    pages = "5558--5561",
    year = "1995"
}

@article{Gundlach:1996eg,
    author = "Gundlach, Carsten",
    title = "{Understanding critical collapse of a scalar field}",
    eprint = "gr-qc/9604019",
    archivePrefix = "arXiv",
    reportNumber = "LAEFF-96-07",
    doi = "10.1103/PhysRevD.55.695",
    journal = "Phys. Rev. D",
    volume = "55",
    pages = "695--713",
    year = "1997"
}

@article{Baumgarte:2023tdh,
    author = {Baumgarte, Thomas W. and Br{\"u}gmann, Bernd and Cors, Daniela and Gundlach, Carsten and Hilditch, David and Khirnov, Anton and Ledvinka, Tom{\'a}{\v{s}} and Renkhoff, Sarah and Fern{\'a}ndez, Isabel Su{\'a}rez},
    title = "{Critical Phenomena in the Collapse of Gravitational Waves}",
    eprint = "2305.17171",
    archivePrefix = "arXiv",
    primaryClass = "gr-qc",
    doi = "10.1103/PhysRevLett.131.181401",
    journal = "Phys. Rev. Lett.",
    volume = "131",
    number = "18",
    pages = "181401",
    year = "2023"
}

@article{Baumgarte:2015dya,
    author = {Baumgarte, Thomas W. and Montero, Pedro J. and M{\"u}ller, Ewald},
    title = "{Numerical Relativity in Spherical Polar Coordinates: Off-center Simulations}",
    eprint = "1501.05259",
    archivePrefix = "arXiv",
    primaryClass = "gr-qc",
    doi = "10.1103/PhysRevD.91.064035",
    journal = "Phys. Rev. D",
    volume = "91",
    number = "6",
    pages = "064035",
    year = "2015"
}

@article{Ruchlin:2017com,
    author = "Ruchlin, Ian and Etienne, Zachariah B. and Baumgarte, Thomas W.",
    title = "{SENR/NRPy+: Numerical Relativity in Singular Curvilinear Coordinate Systems}",
    eprint = "1712.07658",
    archivePrefix = "arXiv",
    primaryClass = "gr-qc",
    doi = "10.1103/PhysRevD.97.064036",
    journal = "Phys. Rev. D",
    volume = "97",
    number = "6",
    pages = "064036",
    year = "2018"
}

@article{Werneck:2021kch,
    author = "Werneck, Leonardo R. and Etienne, Zachariah B. and Abdalla, Elcio and Cuadros-Melgar, Bertha and Pellicer, C. E.",
    title = "{NRPyCritCol {\&} SFcollapse1D: an open-source, user-friendly toolkit to study critical phenomena}",
    eprint = "2106.06553",
    archivePrefix = "arXiv",
    primaryClass = "gr-qc",
    doi = "10.1088/1361-6382/ac33bf",
    journal = "Class. Quant. Grav.",
    volume = "38",
    number = "24",
    pages = "245005",
    year = "2021"
}

@software{engrenage_2026,
  author  = {GRTL Collaboration},
  title   = {{Engrenage}: A Spherically Symmetric {BSSN} Code in Python},
  year    = {2026},
  url     = {https://github.com/GRTLCollaboration/engrenage},
  urldate = {2026-09-30},
  note    = {Software developed by the GRTL Collaboration}
}

@article{Baumgarte:2023saw,
    author = "Baumgarte, Thomas W. and Gundlach, Carsten and Hilditch, David",
    title = "{Critical phenomena in the collapse of quadrupolar and hexadecapolar gravitational waves}",
    eprint = "2303.05530",
    archivePrefix = "arXiv",
    primaryClass = "gr-qc",
    doi = "10.1103/PhysRevD.107.084012",
    journal = "Phys. Rev. D",
    volume = "107",
    number = "8",
    pages = "084012",
    year = "2023"
}

@article{Fernandez:2022hyx,
    author = "Fern{\'a}ndez, Isabel Su{\'a}rez and Renkhoff, Sarah and Cors, Daniela and Bruegmann, Bernd and Hilditch, David",
    title = "{Evolution of Brill waves with an adaptive pseudospectral method}",
    eprint = "2205.04379",
    archivePrefix = "arXiv",
    primaryClass = "gr-qc",
    doi = "10.1103/PhysRevD.106.024036",
    journal = "Phys. Rev. D",
    volume = "106",
    number = "2",
    pages = "024036",
    year = "2022"
}

@article{Mendoza:2021nwq,
    author = "Mendoza, Maria F. Perez and Baumgarte, Thomas W.",
    title = "{Critical phenomena in the gravitational collapse of electromagnetic dipole and quadrupole waves}",
    eprint = "2104.03980",
    archivePrefix = "arXiv",
    primaryClass = "gr-qc",
    doi = "10.1103/PhysRevD.103.124048",
    journal = "Phys. Rev. D",
    volume = "103",
    number = "12",
    pages = "124048",
    year = "2021"
}

@article{Ledvinka:2021rve,
    author = "Ledvinka, Tom{\'a}{\v{s}} and Khirnov, Anton",
    title = "{Universality of Curvature Invariants in Critical Vacuum Gravitational Collapse}",
    eprint = "2102.09579",
    archivePrefix = "arXiv",
    primaryClass = "gr-qc",
    doi = "10.1103/PhysRevLett.127.011104",
    journal = "Phys. Rev. Lett.",
    volume = "127",
    number = "1",
    pages = "011104",
    year = "2021"
}

@article{Baumgarte:2019fai,
    author = "Baumgarte, Thomas W. and Gundlach, Carsten and Hilditch, David",
    title = "{Critical phenomena in the gravitational collapse of electromagnetic waves}",
    eprint = "1909.00850",
    archivePrefix = "arXiv",
    primaryClass = "gr-qc",
    doi = "10.1103/PhysRevLett.123.171103",
    journal = "Phys. Rev. Lett.",
    volume = "123",
    number = "17",
    pages = "171103",
    year = "2019"
}

@article{East:2025nfb,
    author = "East, William E.",
    title = "{Gravitational Collapse in the Vicinity of the Extremal Black Hole Critical Point}",
    eprint = "2511.20567",
    archivePrefix = "arXiv",
    primaryClass = "gr-qc",
    doi = "10.1103/2myd-cfwh",
    journal = "Phys. Rev. Lett.",
    volume = "136",
    number = "15",
    pages = "151401",
    year = "2026"
}

@article{Crump:2026kgu,
    author = "Crump, John R. V. and Gadioux, Maxime and Reall, Harvey S. and Santos, Jorge E.",
    title = "{Violation of the Third Law of Black Hole Mechanics in Vacuum Gravity}",
    eprint = "2601.20955",
    archivePrefix = "arXiv",
    primaryClass = "gr-qc",
    doi = "10.1103/gbg1-pjgq",
    journal = "Phys. Rev. Lett.",
    volume = "136",
    number = "17",
    pages = "171405",
    year = "2026"
}

@article{LIGOScientific:2025rsn,
    author = "Abac, A. G. and others",
    collaboration = "LIGO Scientific, VIRGO, KAGRA",
    title = "{GW231123: A Binary Black Hole Merger with Total Mass 190{\textendash}265 M$_{\odot}$}",
    eprint = "2507.08219",
    archivePrefix = "arXiv",
    primaryClass = "astro-ph.HE",
    reportNumber = "DCC: P2500026-v6, DCC: P2500026-v8",
    doi = "10.3847/2041-8213/ae0c9c",
    journal = "Astrophys. J. Lett.",
    volume = "993",
    number = "1",
    pages = "L25",
    year = "2025"
}

\end{document}